# Crop Yield Time-Series Data Prediction Based on Multiple Hybrid Machine Learning Models

**Yueru Yan[1,4], Yue Wang [1,5], Jialin Li [1,6], Jingwei Zhang[2,7], Xingye Mo[3,8]**

[1] Qufu Normal University, Rizhao, China
[2] SAP Labs China, Shanghai, China
[3] Department of Electrical and Computer Engineering, New York University, New York, USA

[4] 13969497808@163.com
[5] 19558780124@163.com
[6] Lijialin55@outlook.com
[7] rocky.zhang@sap.com
[8] xm2100@nyu.edu

**Abstract.** Agriculture plays a crucial role in the global economy and social stability, and accurate crop yield prediction is essential for rational planting planning and decision-making. This study focuses on crop yield Time-Series Data prediction. Considering the crucial significance of agriculture in the global economy and social stability and the importance of accurate crop yield prediction for rational planting planning and decision-making, this research uses a dataset containing multiple crops, multiple regions, and data over many years to deeply explore the relationships between climatic factors (average rainfall, average temperature) and agricultural inputs (pesticide usage) and crop yield. Multiple hybrid machine learning models such as Linear Regression, Random Forest, Gradient Boost, XGBoost, KNN, Decision Tree, and Bagging Regressor are adopted for yield prediction. After evaluation, it is found that the Random Forest and Bagging Regressor models perform excellently in predicting crop yield with high accuracy and low error. As agricultural data becomes increasingly rich and time-series prediction techniques continue to evolve, the results of this study contribute to advancing the practical application of crop yield prediction in agricultural production management. The integration of time-series analysis allows for more dynamic, data-driven decision-making, enhancing the accuracy and reliability of crop yield forecasts over time.

## 1. Introduction

Agriculture plays a vital role in global economic development and social stability, especially in meeting the growing population's increasing food demands. The stability and growth of crop yield are crucial in this regard. With the advancement of technology, machine learning techniques have been increasingly applied in the agricultural field, providing novel methods and approaches for crop yield prediction. Accurate crop yield prediction not only helps farmers to plan planting schedules and optimize resource allocation effectively but also provides scientific support for governments to formulate agricultural policies.

Time-series analysis, as an important branch of prediction techniques, holds significant potential in agriculture. By leveraging time-series analysis, it is possible to uncover the dynamic relationships between climate changes, agricultural inputs, and crop yield variations, identifying trends, cyclical fluctuations, and critical influencing factors. In modern agricultural production, the impacts of climate factors (e.g., rainfall, temperature, humidity) and agricultural inputs (e.g., fertilizers, irrigation) on crop yields exhibit complex nonlinear patterns, which are often challenging for single prediction models to capture comprehensively.

This study utilizes relevant datasets and integrates multiple machine learning algorithms, including Linear Regression, Random Forest, Gradient Boost, XGBoost, K-Nearest Neighbors (KNN), Decision Tree, and Bagging Regressor, to analyze the dynamic effects of climate factors and agricultural inputs on crop yields. By combining these algorithms, the study aims to capture nonlinear characteristics and time-series trends more effectively, thereby improving prediction accuracy and stability. The findings provide valuable decision-making support for agricultural production and serve as an important reference for addressing uncertainties in future agricultural practices.

## 2. Literature review

In the agricultural field, crop yield prediction is crucial for rational planning of planting, optimizing resource allocation, and formulating agricultural policies. In recent years, numerous studies have been dedicated to improving the accuracy and reliability of crop yield prediction using machine learning techniques. The following is a review of the relevant references.

Schwalbert et al. integrated machine learning and weather data for satellite-based soybean yield forecasting in southern Brazil, demonstrating the effectiveness of multi-source data fusion for regional crop yield prediction. Their study emphasizes using satellite data to monitor crop growth and analyze meteorological factors **[1]**. Paudel et al. highlighted the utility of machine learning in large-scale crop yield forecasting, providing practical guidance for national and regional food production estimates and agricultural resource allocation **[2]**.

Bharadiya et al. comprehensively used remote sensing data, agrarian factors, and machine learning methods for crop yield prediction. This study demonstrates the advantages of multi-dimensional data integration. Remote sensing data can provide real-time information on the growth of large areas of crops, and combined with traditional agricultural factors, it enables the prediction model to more comprehensively capture various factors affecting crop yield, thereby improving prediction accuracy **[3]**.

Mishra et al. reviewed the applications of machine learning techniques in agricultural crop production, summarizing the research status and development trends in this field at that time. This review provides comprehensive background information for subsequent studies and clarifies the main directions of machine learning applications in agriculture, including crop yield prediction, pest and disease monitoring, soil quality assessment, and other aspects **[4]**.

Filippi et al. adopted a method that combines multi-layered, multi-farm datasets with machine learning to predict grain crop yields, emphasizing the importance of data stratification and diversification for improving the generalization ability of prediction models. By integrating multi-source data from different farms, the model can learn a wider range of crop growth patterns and environmental adaptability, providing a more targeted and practical method for yield prediction in precision agriculture **[5]**.

Johnson et al. predicted crop yields on the Canadian Prairies using vegetation indices and machine learning, showcasing remote sensing's advantages in monitoring large-scale crop growth. Their approach provides valuable insights for agricultural management in prairie regions **[6]**. Despite progress in crop yield prediction, challenges remain in integrating diverse factors, improving model accuracy in complex environments, and developing practical decision-support tools. This study leverages a time-series dataset spanning multiple crops, regions, and countries to analyze variables like climate and agricultural inputs. A hybrid multi-machine learning model is proposed to address the complexities of multi-dimensional time-series data prediction.

## 3. Data

*3.1. Data collection*

The dataset used in this study contains relevant information on multiple crops in various countries and regions from 1990 to 2013. The variables in the dataset include crop yield (hg/ha_yield), average rainfall (average_rain_fall_mm_per_year), pesticide usage (pesticides_tonnes), average temperature (avg_temp), as well as region (Area), crop type (Item), and year (Year). These data provide a rich source of information for analyzing the impact of climatic factors and agricultural inputs on crop yield.

*3.2. Descriptive statistical analysis*

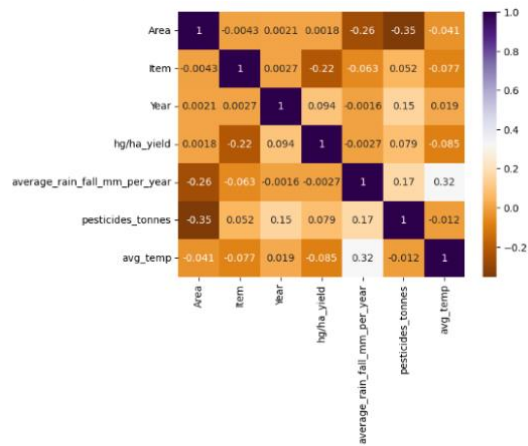

**Figure 1.** Correlation heat map

This is a correlation matrix diagram showing the correlations among multiple variables. The variables in the diagram include Area, Item, Year, hg/ha_yield (crop yield), average_rain_fall_mm_per_year (average rainfall), pesticides_tonnes (pesticide usage), and avg_temp (average temperature). The colors range from dark purple to light orange, indicating correlations from negative to positive, with darker colors representing stronger correlations. From the diagram, it can be seen that the crop yield (hg/ha_yield) is negatively correlated with the average rainfall (average_rain_fall_mm_per_year) and positively correlated with the pesticide usage (pesticides_tonnes). The average rainfall is positively correlated with the pesticide usage, and the average temperature (avg_temp) is also positively correlated with the average rainfall. The correlations among other variables are relatively weak. This matrix is helpful for analyzing the direction and degree of the influence of different factors on crop yield.

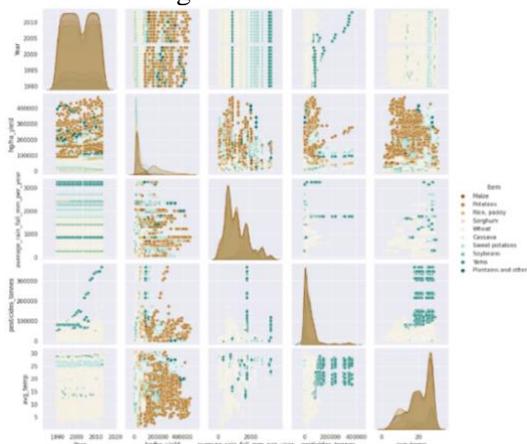

**Figure 2.** Descriptive graph

Yams are prone to pests and diseases, leading to significant pesticide use. They adapt well to varying rainfall conditions due to their deep root systems, thriving in both humid and arid environments. However, excessive or insufficient rainfall can impact their yield and quality, suggesting future research on rainfall's effect on yam productivity for better agricultural management.

Wheat demonstrates wide temperature adaptability and can thrive in low-rainfall areas thanks to its efficient water use, deep roots, and water-retaining structures. However, successful cultivation in such conditions requires proper management, including irrigation and soil moisture retention. Further studies could enhance wheat productivity in arid regions.

Sorghum and soybeans yield lower than other crops due to specific challenges. Sorghum is sensitive to soil fertility, water availability, and sunlight, while soybeans struggle in nitrogen-deficient soils due to limited nitrogen-fixation abilities. Addressing these limitations could improve their yields.

Potatoes are highly productive crops, known for their adaptability, efficient photosynthesis, strong tuber formation, and resilience to environmental changes. Research into their productivity mechanisms may provide insights to enhance yields of crops like sorghum and soybeans, contributing to global food security. Prioritizing soil fertility optimization, improved planting techniques, and breeding high-yield varieties could significantly boost agricultural output.

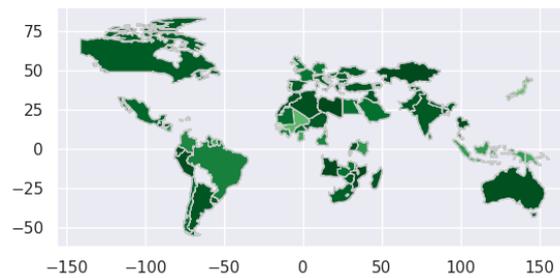

**Figure 3.** Data distribution map

This is a world map with different shades of green representing different values. The color scale on the right side ranges from -150 to 150, with corresponding values marked at intervals of 25. Each country or region is colored according to its respective value within this range. The map provides a visual representation of how these values are distributed across the globe, allowing for easy comparison between different geographical areas.

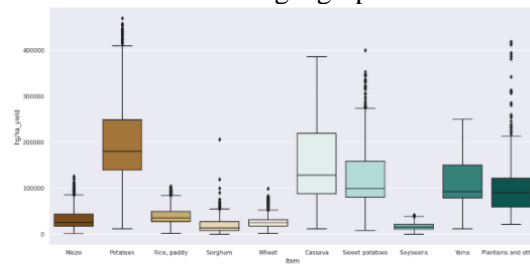

**Figure 4.** Boxplot

Figure illustrates the yield of various agricultural items. The y - axis represents the yield values, ranging from 0 to 45000, while the x - axis lists the different items.Maize has the highest yield among the items shown, with a value significantly above 40000. Potatoes follow with a yield around 25000. Other items such as Rice, paddy, Sorghum, Wheat, Cassava, Sweet potatoes, Soybeans, Yams, and Plantains and others have relatively lower yields, with some values being less than 5000.Each figure color-coded differently, indicating different categories of crops.

**4. Model introduction**

This study employs several machine learning models for crop yield prediction, including Linear Regression, Random Forest, Gradient Boost, XGBoost, KNN, Decision Tree, and Bagging Regressor.

Linear Regression assumes a linear relationship between variables and minimizes the squared errors between predicted and actual values. It is computationally efficient but struggles with complex nonlinear relationships. Random Forest, an ensemble method based on decision trees, improves model accuracy by aggregating multiple tree predictions. It handles high-dimensional data well, has strong anti-overfitting capabilities, and is robust to noise and outliers [7].

Gradient Boost builds decision trees iteratively, optimizing the model based on previous residuals, improving predictive performance with each step. While effective for complex relationships, it is prone to overfitting and requires careful parameter tuning. XGBoost, an enhanced version of Gradient Boost, offers faster computation, better performance, and incorporates regularization techniques to prevent overfitting. It excels with large datasets due to its support for parallel computing [8].

The KNN algorithm predicts based on the distance between samples, using the K nearest neighbors in the training set. It is simple and intuitive but computationally expensive and sensitive to the data's local structure. The Decision Tree model constructs a tree-like structure to make predictions, offering interpretability but being prone to overfitting, especially with complex or small datasets. Bagging Regressor, an ensemble method, improves stability and generalization by averaging predictions from multiple base learners, reducing variance and enhancing model robustness.

## 5. Model analysis and discussion

**Table 1. Model results analysis**

| Model | Accuracy | MAE | MAPE | R2 |
|---|---|---|---|---|
| Linear Regression | 0.074 | 6095.32 | 2.419 | 0.074 |
| Random Forest | 0.986 | 348.84 | 0.103 | 0.986 |
| Gradient Boost | 0.831 | 2118.66 | 0.597 | 0.831 |
| XGBoost | 0.973 | 734.95 | 0.209 | 0.973 |
| KNN | 0.288 | 4771.36 | 1.631 | 0.288 |
| Decision Tree | 0.976 | 355.27 | 0.096 | 0.976 |
| Bagging Regressor | 0.986 | 345.51 | 0.101 | 0.986 |

The results in table 1 show that the linear regression model performs poorly, with an accuracy of 0.074 and a high MAE, indicating its inability to effectively predict crop yield due to the complex, nonlinear nature of the data. In contrast, the Random Forest and Bagging Regressor models achieve high accuracy (0.986) and low MAEs, demonstrating their ability to capture intricate data patterns. Random Forest improves generalization by using multiple decision trees, while Bagging Regressor enhances stability through aggregation of predictions. The Gradient Boost and XGBoost models also yield high accuracy (0.831 and 0.973, respectively), but their slightly higher MAEs suggest some overfitting when handling complex relationships. These findings highlight the strengths and weaknesses of various models in crop yield prediction.

The KNN model has a relatively low accuracy of only 0.288, and relatively large MAE. This may be because the KNN model is too sensitive to the local structure of the data and is easily affected by noise and outliers when dealing with large-scale and high-dimensional data, resulting in poor prediction performance [9].

The Decision Tree model has a relatively high accuracy of 0.976, but its stability may be slightly worse compared with the Random Forest and Bagging Regressor models. The decision tree is prone to overfitting, especially when the data is complex or the sample size is limited, and it may over-learn the details in the training data, thereby affecting the prediction ability for new data [10].

In summary, the Random Forest and Bagging Regressor models perform the best in this study and are suitable for predicting crop yield.

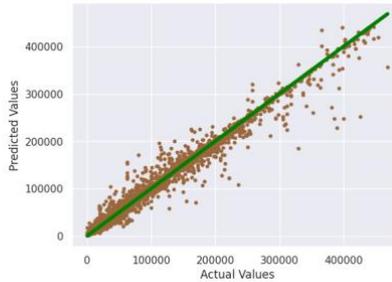

**Figure 5.** Bagging Regressor model results

Figure 5 compares the predicted values from the Bagging Regressor model with actual crop yields. The trend of the predicted and actual values aligns closely, with most data points near the diagonal, indicating the model's strong ability to fit the data and make accurate predictions. With an accuracy of 0.986 and low MSE and MAE, the model proves effective for crop yield prediction. This model can help farmers optimize planting strategies and inputs, such as pesticide use, while aiding government agencies in formulating policies and allocating resources, particularly in regions with low predicted yields.

**6. Conclusions**

This study explores the relationships between climatic factors (rainfall, temperature) and agricultural inputs (pesticide usage) with crop yield, constructing multiple machine learning models for prediction. Evaluation reveals that the Random Forest and Bagging Regressor models exhibit high accuracy and stability, effectively leveraging climatic and agricultural data for yield prediction. These findings offer valuable insights for agricultural decision-making. Farmers can optimize planting plans and pesticide usage to improve yield and quality, while governments can use the models for policy formulation and resource allocation, such as providing irrigation in low-yield areas. Future research could incorporate additional factors, like soil quality and planting density, and explore advanced machine learning techniques to enhance prediction accuracy and support sustainable agricultural development.